\documentclass[%
 aip,
 amsmath,amssymb,
 reprint,%
]{revtex4-1}

\usepackage{graphicx}
\usepackage{dcolumn}
\usepackage{bm}

\usepackage[utf8]{inputenc}
\usepackage[T1]{fontenc}
\usepackage{mathptmx}
\usepackage{etoolbox}
\usepackage{color}
\usepackage{soul}

\makeatletter
\def\@email#1#2{%
 \endgroup
 \patchcmd{\titleblock@produce}
  {\frontmatter@RRAPformat}
  {\frontmatter@RRAPformat{\produce@RRAP{*#1\href{mailto:#2}{#2}}}\frontmatter@RRAPformat}
  {}{}
}%
\makeatother
\begin{document}

\preprint{AIP/123-QED}

\title[]{RANDOM MULTI-PLAYER GAMES}


\author{Natalia L. Kontorovsky}
\email{natilkontorovsky@gmail.com}
\affiliation{Instituto de C\'{a}lculo, FCEN, Universidad de Buenos Aires and CONICET, 
 Av Cantilo s/n, Ciudad Universitaria
 (1428)  Buenos Aires, Argentina}  

\author{Juan Pablo Pinasco}%
 \email{jpinasco@gmail.com}
 \homepage{https://mate.dm.uba.ar/~jpinasco/}
\affiliation{Departamento  de Matem{\'a}tica and IMAS UBA-CONICET,
 Facultad de Ciencias Exactas y Naturales, Universidad de Buenos Aires,
 Av Cantilo s/n, Ciudad Universitaria
 (1428) Buenos Aires, Argentina.}

\author{Federico Vazquez} 
\email{fede.vazmin@gmail.com}\homepage{https://fedevazmin.wordpress.com}
\affiliation{Instituto de C\'{a}lculo, FCEN, Universidad de Buenos Aires and CONICET, 
 Av Cantilo s/n, Ciudad Universitaria
 (1428) Buenos Aires, Argentina}

\date{\today}

\begin{abstract}
The study of evolutionary games with pairwise local interactions has been of interest to many different disciplines. Also local interactions with multiple opponents had been considered, although always for a fixed   amount of players. In many situations, however, interactions between different numbers of players in each round could take place, and this   case can not be reduced to pairwise interactions. In this work we formalize and generalize the definition of evolutionary stable strategy (ESS) to be able to include a scenario in which  the game is played by two players with probability $p$, and by three players with the complementary probability $1-p$. We show the existence of equilibria in pure and mixed strategies depending on the probability $p$, on a concrete example of the duel-truel game.
We find a range of $p$ values for which the game has a mixed equilibrium and the proportion of players in each strategy depends on the particular value of $p$. We prove that each of these mixed equilibrium points are ESS. A more realistic way to study this  dynamics with high-order interactions is to look at how it evolves in complex networks.
We introduce and study an agent-based model  on a network with a fixed number of nodes, which evolves as the replicator equation predicts. By studying the dynamics of this model  on random  networks we  find that the phase transitions between the pure and mixed equilibria depend on the probability $p$ and also on the mean degree of the network.
\end{abstract}

\maketitle

\begin{quotation}

Game theory has had remarkable success as a framework to study the behavior of large populations where individuals playing different strategies interact through some game, and they can replicate according to their payoffs. This theory emerged to create a biological context where evolution could be reflected. Recently, however, evolutionary game theory has become of increased interest to economists, sociologists, and anthropologists and social scientists in general. 
It suggested new interpretations and prompted new observational
studies. This led us to study the different ways in which populations can interact. Most studies have considered populations where individuals compete in pairwise interactions or even more than two (possibly many) individuals compete simultaneously, but always with a fixed amount of players. However, more complicated scenarios were less investigated.
The present paper studies a model that incorporates interactions in symmetric games with random number of players, which allowed us to find games where the coexistence of strategies occurs and is a stable state in time.
We give an insight into the conditions to obtain this coexistence states and investigate by many different approaches, what parameters of the game or the environment they depend on.

\end{quotation}

\section{\label{sec:level1}INTRODUCTION\protect }

Evolutionary game theory provides a framework to study the behavior of large populations where individuals playing different strategies (or having different biological traits) interact through some game, and they can replicate according to their payoffs.  Classically, it is assumed that each individual is equally likely to face any other
individual, and a mean field approximation gives a system of ordinary differential equations for the time evolution of the proportion of individuals in each strategy.

The concept of evolutionary stable strategy (ESS) was
introduced by Maynard Smith and Price\cite{smith1973logic}, and it can be characterized in terms of the payoffs among members
of the original population (incumbents) and the invaders (or mutants). We can understand it
as a refinement of Nash equilibria since implies a collective no-deviation condition at least
up to a critical fraction of players, usually known as the invasion barrier, instead of considering single-player deviations. Hence, an ESS satisfies the additional, stronger condition of stability, which implies that if an ESS is reached, then the proportions of players playing the different strategies do not change over time\cite{hines1987evolutionary,hofbauer2003evolutionary,riley1979evolutionary,smith1982evolution}. 

Usually, evolutionary games with pairwise interactions were studied\cite{pinasco,smith1986evolutionary,zeeman1980population}. Also, local interactions with various opponents had been considered, but in each round the same fixed number of players are randomly selected from the population to play against each other\cite{gokhale2010evolutionary,Multi1997}.
On the other hand, there are many situations in which the number of players can vary over time and even between rounds. It also happens that the optimal strategies in a two players game could not be optimal in a three players game, so interactions between multiple players can not be reduced to pairwise interactions. Therefore, it is interesting to model and study the case in which
the game can be played by a different number of players in each round when the strategy must be selected previously, without knowing a priori the exact number of players involved.

In this article we formalize and generalize the definition of evolutionary stable strategy to be able to include this scenario. 
This  question was analyzed previously\cite{tembine2008evolutionary}, however only  two combinations of incumbents and mutants were considered. 
As we show in \S~\ref{EGT}, all the combinations must  be considered, and a hierarchy of payoffs is needed in order to characterize an ESS when the number of players in each interaction is a random variable.

 In order to explore these questions we study here the simplest non-trivial case of the duel-truel game. As usual, in a duel two players aim to  eliminate each other, while in a truel three players are involved. Truel games were introduced independently by 
Shubik\cite{shubik1982game} and Epstein\cite{epstein2012theory}, and their mechanics is as follows.  At each time step, an order is drawn between the three players and they shoot cyclically deciding  who their target will be (this is usually expressed as the players {\it shooting} and {\it killing} each other, although
possible applications of this simple game do not need to be so violent).   Each player aims at its chosen opponent and
with a given probability it  achieves the goal of eliminating this player from the game. This process repeats until only one of the three players remains.

The paradox is that the player that has the highest
probability of annihilating competitors does not need to be necessarily the winner of
this game. This surprising result was already present in the early literature on truels\cite{amengual2006truels,shubik1982game,kilgour1975sequential,toral2005distribution}. The case where this disparity is most clearly seen, in which the one with the least probability of killing turns out to be the one with the highest probability of surviving, is when they are allowed to shoot into the air. Thinking about a round, in the event that two players have perfect aim and one does not, if the player with the worst aim has to shoot first, it will prefer to shoot into the air, instead of shooting at one of the opponents, since if they kill one of them, it will be killed by the remaining player. Thus, the perfect ones shoot each other, which results in that the survival probability of the player with the worst aim increases.

For simplicity, we will here consider only two strategies, two classes of players with different probabilities of killing  the chosen opponent, {\it mediocre} players which have   a killing probability equal to $1/2$, and {\it perfect} players which have a killing probability equal to $1$. The payments of the matrix in the duel game favor the population that uses the strongest strategy, while the payments of the matrix of the truel game favor the population that pays the weakest strategy. In other words, if the local interactions between the players in the population would be pairwise, the ESS would be for the entire population to play strategy {\it perfect}, whereas if local interactions involve three players, the ESS would be for the entire population to play strategy {\it mediocre}.

This led us to consider what the ESS would be in a scenario where the number of players is a random variable, the so-called Poisson games\cite{myerson1998population}. As an example, agents will play a duel with probability $p\in(0,1)$, and a truel with probability $1-p$.
Let us mention that this game was considered previously \cite{archetti2012survival}, and only the survival probabilities of each strategy were obtained without considering the optimal option of shooting into the air.

Also,  we introduce an agent-based model in which players interact in a complex network by copying the strategies of their neighbors, with a dynamics that  in mean-field evolves following the replicator equation. Let us observe that the replicator dynamics is usually defined in terms of new individuals entering the population, by selecting  a pure strategy with a probability proportional to the payoff given the current mix of agents. Some related works have also explored similar issues \cite{traulsen2005coevolutionary}, but they  assume that individuals reproduce instead of copying strategies.
 
Our approach has an independent interest, and has the advantage that can be used in networks  with a fixed number of nodes or agents, bypassing the issue of how to add new nodes as agents replicate.

We perform extensive Monte Carlo (MC) simulations of the model in different types of networks, and develop an analytical approach based on a pair approximation \cite{Vazquez-2008-1,Vazquez-2008-2,Vazquez-2010,Demirel-2014} that allows to obtain approximate equations for the evolution of the fraction of perfect agents in the network.  This approach enable us to investigate if the transitions between ESS in pure and mixed strategies found within the Nash equilibrium theory are also observed in complex networks, and to identify how the networks' topology affects the existence of mixed equilibria.

The article is organized as follows. In Section \S~\ref{EGT} we introduce some notations and definitions from game theory.  We characterize the ESS in Section \S~\ref{two-strategies} and describe the particular case of the Duel-Truel game in Section \S~\ref{duel-or-truel}. In \S~\ref{model} we introduce the agent-based model, develop a mean-field and a pair approximation approach for the system's dynamics, and compare the results with those of MC simulations of the model in various complex networks. Finally, in Section \S~\ref{conclusions} we give a summary and conclusions.

\section{Evolutionary game theory}
\label{EGT}

Let $G=(S,U,\{1,...,N\})$ be a finite game, where $S$ is the set of  pure strategies $s_1,\ldots, s_m$,  $\{1,...,N\}$ is the set of players, and $U:S^N\to \mathbb{R}^N$ is the payoff of the game when each player selects a strategy.

Let us consider the simplex of mixed strategies, the set of probability vectors, $$\Delta_S=\Big\{x=(x_1,\ldots,x_m) 
: 0\le x_i, \sum x_i=1\Big\},$$ 
where a player using the mixed strategy $x\in \Delta_S$ plays the pure strategy $s_i$ with probability $x_i$.

We extend the payoff $U$ to the set of mixed strategies, in terms of the expected payoffs $E[x^1,\ldots, x^N]$, that is, the sum over the possible payoffs multiplied by the probability they occur, where $x^j$ is the  mixed strategy of player $j$ for $1\le j\le N$.

We say that a Nash equilibrium is a vector of mixed
strategies, one for each player, such that no player can improve its payoff by changing its strategy\cite{hofbauer2003evolutionary}.  Moreover,  $x$ is a symmetric Nash equilibrium if 
$$
E[x,x,\ldots,x]\ge E[y,x,\ldots, x] ~~~ \mbox{ for $y \ne x$.}
$$
On the other hand, let us observe that this definition is restricted only to players involved in a game. The notion of ESS hold for populations of players interacting through some dynamics involving a game,  as we describe below.

  Given a two players
game with finitely many pure strategies $s_1,\ldots, s_m$,
 let us suppose that the proportion of agents in a population playing strategy $s_i$ is $x_i$, so the agents define a mixed strategy $x=(x_i,...,x_m)$. 

Now, the evolution of  agents playing $s_i$ is given by  the following system of 
ordinary differential equations:
$$
\frac{d}{dt}x_i(t)=x_i \big\{ E[x_i,x]-E[x,x] \big\},
$$
where $1\le i\le m$, which is known as the {\it replicator equation}. 
We say that the population distribution  $x$ is an {\it evolutionary stable strategy} if $x$ is a local stable equilibrium of the  dynamics.

Usually, the stability is formulated as first and second order conditions in terms of the incumbent population playing $x$, and any other mutant population distributed in the strategies according to
a proportion $y$:
\begin{equation}\label{a1}
    E[x,x]> E[x,y]
\end{equation} 
\textit{or}
\begin{eqnarray}\label{a2}
    E[x,x] &=& E[x,y], \\
    E[x,x] &>& E[y,y].
\end{eqnarray}
Since the condition $E[x,x]\ge E[x,y]$ holds because $x$ is a Nash equilibrium, we get that an incumbent strictly outplay a mutant, or get an advantage whenever it plays with another incumbent, while mutants play among them.

\bigskip
Our next task is to extend this definition to symmetric games with random number of players.  For that, we consider a population of agents that play a given game.  With certain probability $p$ two players are chosen at random and play the game against each other, and with the complementary probability $1-p$ three players are chosen at random and play the game among themselves.

We need to distinguish between
the expected payoff $E[x_i,x]$ of an agent playing $x_i$ against a population
playing the different strategies with a given proportion $x$ in the two players game, and $E_G[x_i,x]$, the expected payoff in the game with a random number of opponents, which in
our case is
$$
E_G[x_i,x] = p E[x_i,x] + (1-p)E[x_i,x,x].
$$
Hence we get the system of ordinary differential equations
 $$
\frac{d}{dt}x_i(t)=x_i \big\{ E_G[x_i,x]-E_G[x,x] \big\}.
$$ 

Let us assume that the strategy
$x$ is played by the incumbents,   while a small fraction $\varepsilon$ of mutants  
plays $y$.  We say that $x$ is \textit{evolutionary
stable} (ES) against $y$ if there exists
$\varepsilon_0>0$ such that
\begin{align}
  & p \Big\{ \varepsilon E[x,y]+ (1-\varepsilon)E[x,x] \Big\} + \nonumber \\ 
&     (1-p) \Big\{ 2 \varepsilon(1-\varepsilon) E[x,y,x] 
    +(1-\varepsilon)^2 E[x,x,x]+\varepsilon ^2 E[x,y,y] \Big\} \nonumber
   \\
   &>     \label{condition}  \\ 
 &   p  \Big\{ \varepsilon E[y,y]+ (1-\varepsilon)E[y,x] \Big\} +
    \nonumber \\ 
        & (1-p) \Big\{ 2 \varepsilon(1-\varepsilon) E[y,y,x]
    +(1-\varepsilon)^2 E[y,x,x]+\varepsilon^2 E[y,y,y] \Big\}
 \nonumber
\end{align} 
for any $\varepsilon\le \varepsilon_0$.

We say that $x$ is an {\it evolutionary stable strategy} (ESS)
if there exists
$\varepsilon_0>0$ such that   $x$ is ES against $y$ for any $y\neq x$.

For convenience, let us call $d_{i,j}$ the difference between the expected
payoffs of and agent playing $x$ and another agent playing $y$, when both play against a number of $i$ and $j$ opponents that play $x$ and $y$, respectively.

Thus, 
\begin{align*}
&d_{1,0} = E[x,x]-E[y,x], \\
&d_{2,0} = E[x,x,x]-E[y,x,x], \\
&d_{0,1} = E[x,y] - E[y,y], \\
&d_{0,2} = E[x,y,y]- E[y,y,y], \\
&d_{1,1} = E[x,y,x] - E[y,y,x]. 
\end{align*}
Then, the condition from Eq.~(\ref{condition}) is equivalent to the following: a strategy $x$ is ES against a strategy $y$ if either 
\begin{align}\label{desig1}
p\ d_{1,0} + (1-p)d_{2,0}&>0, 
\end{align}
\textit{or} 
\begin{align}\label{desig2}
p \ (d_{0,1} - d_{1,0}) + (1-p)2(d_{1,1}-d_{2,0})&>0 \\   p \ d_{1,0} + (1-p)d_{2,0}&=0,
\end{align} 
\textit{or} \begin{align}\label{desig3}
(1-p)(-2d_{1,1}+d_{2,0}+d_{0,2})& >0 \\
   p \ (d_{0,1} - d_{1,0}) + (1-p)2(d_{1,1}-d_{2,0})&=0 \\
  p \ d_{1,0} + (1-p)d_{2,0} &=0.
\end{align}
We call \eqref{desig1} label $0$ condition, \eqref{desig2} label $1$ condition and \eqref{desig3} label 2 condition. The level is given by the power of $\varepsilon$ that we are looking at in each case.

It easy to generalize this idea to games where the number of players can be any integer whenever the same set of finitely many strategies  is used.
 In the characterization of ESS analyzed previously\cite{tembine2008evolutionary} only two cases of interactions were considered, incumbents against an incumbent vs. incumbents
against a mutant, or intra-group interactions (all
mutants-all incumbent), as in equations \eqref{a1} and \eqref{a2}.

\section{Two strategies games}
\label{two-strategies}

For the sake of simplicity, we consider only two strategies, $s_{1}$ and $s_{2}$, and two symmetric games with these strategies, one for two players and the other for three players.  With probability $p$ a pair of agents is chosen to play a game, and with probability $1-p$ a triplet is chosen to play.  We denote by $a_{ij}$ the payoff of an agent that plays strategy $s_i$ against another agent that plays strategy $s_j$, in a two--player game.  Similarly, $a_{ijk}$ denotes the payoff of an agent playing strategy $s_j$ against other two agents that play strategies $s_j$ and $s_k$, in a three-player game.  Here the subindices $i,j,k$ take the values $1$ or $2$.  Recall that $a_{121} = a_{112}=a_{211}$ and $a_{221} = a_{212}= a_{122}$ due to the symmetry of the game, and also $a_{11}=a_{22}=1/2$ and $a_{111}=a_{222}=1/3$. 

For the two players game, if player $I$ plays $s_i$
and player $II$ plays $s_j$, then $I$ receives the payoff  $a_{ij}$ and $II$ receives $a_{ji}$. The values $a_{ij}$ can be thought as elements of a $2\times 2$ matrix $A_{2}$.  In the same way, if three players $I$, $II$ and $III$ are involved, such that they play strategies $s_i$, $s_j$ and $s_k$, respectively, then $I$ receives the payoff $a_{ijk}$, $II$ receives $a_{jik}$ and $III$ receives $a_{kij}$.  In this case, the values $a_{ijk}$ are considered as the elements of two $2\times 2$ 
matrices $A_{3k}$.  As usual, the matrices $A_{2}$, $A_{31}$ and $A_{32}$ are constant. 

 The set of mixed strategies is $\Delta_S=\{(\sigma,1-\sigma) : \sigma \in [0,1]\}$. The expected payoff of a player using $(\sigma,1-\sigma)$ against an opponent that uses strategy $s_j$ is $\sigma a_{1j} + (1-\sigma)a_{2j}$, and when it plays against two opponents using $s_j$ and $s_k$ is $\sigma a_{1jk} + (1-\sigma)a_{2jk}$.  This way of computing expected payoffs also applies when more than player use mixed strategies.

Since the simplex of strategies is the segment $[0,1]$, we can identify each strategy with a single number, the probability to play $s_{1}$. So, for brevity we write
$E[\sigma,t]$ instead of $E[(\sigma,1-\sigma), (t,1-t)]$.

Let us suppose that a proportion $\sigma(t)$ of the population plays the first strategy at time $t$, and
$1-\sigma(t)$ plays $s_2$. Assuming that the players are selected at random from the population, the 
evolution of this proportion is given by the replicator equation,
$$
\frac{d\sigma}{dt}  =\sigma \big\{ E_G[1,\sigma]-E_G[\sigma,\sigma] \big\},
$$ 
which can be expanded to obtain a more symmetric expression,
$$
\frac{d\sigma}{dt}  = \sigma(1-\sigma) \big\{ E_G[1,\sigma]-E_G[0,\sigma] \big\},
$$

\subsection{Pure ESS.}

Let us consider first the strategy $s_1$, and let $y = (q,1-q)$ be any mixed strategy.
Then from Eq.(\eqref{desig1}), we have 
\begin{eqnarray*}
    p \Big\{ \frac{1}{2}- \left[ q\frac{1}{2}+ (1-q)a_{21} \right] \Big\} &+& \\
    (1-p) \Big\{ \frac{1}{3}- \left[ q\frac{1}{3}+(1-q)a_{211} \right] \Big\} &>& 0,
\end{eqnarray*}
and
\begin{align*}
    &(1-q) \left[ p \left( \frac{1}{2}-a_{21} \right)+ (1-p) \left( \frac{1}{3}-a_{211} \right) \right] >0.
\end{align*}

It follows that the pure strategy $s_1$ is an ESS if
\begin{align}\label{condPURE1}
 p\left(\frac{1}{2}-a_{21}\right)+ (1-p)\left(\frac{1}{3}-a_{211}\right)>0,
\end{align}
whereas the pure strategy $s_{2}$ is an ESS if
\begin{align}\label{condPURE2}
    &p\left(\frac{1}{2}-a_{12}\right)+ (1-p)\left(\frac{1}{3}-a_{122}\right)>0.
\end{align}

\subsection{Mixed ESS.} 

Let $x= (\sigma,1-\sigma)$ be a Nash equilibrium. Then in order to be an ESS, 
$x$ must satisfy $\eqref{desig2}$ or $\eqref{desig3}$
for any  $y = (q,1-q)\neq x$.

Condition \eqref{desig2} is satisfied if
\begin{widetext}
\begin{equation} \label{mixed ESS}
 (\sigma-q)^2 \left( p \left(a_{12}+a_{21}-1\right) + (1-p)2\left((a_{211}-\frac{1}{3})\sigma+(1-\sigma)(a_{122}-\frac{1}{3})+ (2\sigma-1)(a_{121}-a_{221})\right)\right)>0,
\end{equation}
\end{widetext}
while in case of equality, condition \eqref{desig3} must holds
\begin{align}
    & (\sigma-q)\left((\frac{1}{3}-a_{211})+(a_{122}-\frac{1}{3})- 2(a_{121}-a_{221})\right)>0.
\end{align}
Let us note that this condition will always depend on the sign of the term $(\sigma-q)$. This lead us to conclude that $x$ is an ESS if and only if it satisfies condition \eqref{mixed ESS}.

In the general case when $k$ players are involved in the game  with probability $p_k$,
we will  have a similar phenomenon. The inequality 
in the condition for being an ESS for the even labels will depend on the sign of $(\sigma-q)$, then there will be no ESS of those levels.

\section{Duel or Truel game}
\label{duel-or-truel}

We consider now the Duel-Truel random game.  
The Duel and the Truel are both matrix games, and we are going to consider only two strategies $s_{1}$ and $s_{2}$, with strategy $s_1$ the player kills with probability $1$, which we call \textit{perfect} strategy, and the with strategy $s_2$ kills with probability $0.5$, which we call  \textit{mediocre} strategy.

Each player's actions are always taken to improve their chance of survival.
The Duel matrix game is given by: 
$$
\begin{matrix}
 & s_{1} & s_{2} & \\
s_{1} & (1/2,1/2) & (3/4,1/4)\\
s_{2} & (1/4,3/4) & (1/2,1/2)  
\end{matrix},
$$
while the two Truel matrices are given by: 

$III$ plays $s_{1}$:
$$
\begin{matrix}
 & s_{1} & s_{2} & \\
s_{1} & (1/3,1/3,1/3) & (1/4,1/4,1/2) \\
s_{2} &  (1/4,1/2,1/4) & (7/24,17/48,17/48)
\end{matrix}
$$ 

$III$ plays $s_{2}$:
$$
\begin{matrix}
 & s_{1} & s_{2} & \\
s_{1} & (1/2,1/4,1/4) & (17/48,7/24,17/48)\\
s_{2} &  (17/48,17/48,7/24) & (1/3,1/3,1/3)
\end{matrix}
$$
 
Then, we compute the expected payoff of $s_1$ and $s_2$, and get

\begin{eqnarray*}
E[1,\sigma] &=& p\left(\frac{3}{4}-\sigma \frac{1}{4}\right) \\  &+& (1-p)\left[ (1-\sigma)^2\frac{7}{24}+ \sigma(1-\sigma)\frac{1}{2}+ \sigma^2\frac{1}{3} \right], \\
E[0,\sigma] &=& p \left(\frac{1}{2}-\sigma \frac{1}{4}\right) \\ &+& (1-p)\left[ \frac{1}{3}(1-\sigma)^2+\sigma(1-\sigma)\frac{17}{24}+\sigma^2\frac{1}{2} \right]. 
\end{eqnarray*}

Thus, the replicator equation is  
\begin{eqnarray}
\frac{d\sigma}{dt}  = \sigma(1-\sigma) \left[ \frac{p}{4} -  (1-p) \left( \frac{1}{24}+\frac{1}{8}\sigma \right) \right],
\label{replicatoreq}
\end{eqnarray}
which has three fixed points, $\sigma=0$, $\sigma=1$ and 
\begin{equation}
    \sigma^* = \frac{7p - 1}{3(1- p)}.
    \label{sigma*}
\end{equation}
From the Folk Theorem\cite{cressman}, the interior rest point is a Nash equilibrium, however the boundary points can be Nash equilibria or not.

Since $\sigma^{*}$ is the proportion of the population playing strategy $s_{1}$, we need $0<\sigma^{*}<1$, which holds for  $\frac{1}{7}<p<\frac{2}{5}$. Also, by using conditions \eqref{condPURE1} and \eqref{condPURE2}, we get that the Nash equilibrium of the game depends on the value of $p$ that we are using,
$$\sigma_{NE} = 
\left\{ \begin{array}{ccc}
    \ 0 & \mbox{if} & p \leq \frac{1}{7}, \\
    \\
    \ \sigma^{*} & \mbox{if} &\frac{1}{7}<p<\frac{2}{5},\\
    \\
    \ 1 & \mbox{if} & \frac{2}{5} \leq p.
\end{array}\right.
$$
The pure Nash equilibria are ESS. In order to analyze the mixed equilibrium, we need to consider inequality \eqref{desig2}, and this proves that the mixed equilibrium is an ESS for $\frac{1}{7}<p<\frac{2}{5}$.

\bigskip
Let us observe that the killing probabilities $0.5$ and $1$ for the \textit{mediocre} and \textit{perfect} strategies respectively, are not critical for  the results we obtained. A sensitivity analysis shows that we can vary both probabilities in certain
range and the results still hold, changing only the intervals for $p$.

\section{Agent-based dynamics on complex networks}
\label{model}

In this section we introduce and study a model of interacting agents
on a complex network, which provides a kinetic approach to the
duel-truel problem.  The system consists of a population of $N$ agents that
play the duel and truel games, and are allowed to update their
strategies (perfect vs mediocre) as they interact with other agents.
Each agent is located at a node of a complex network of $N$ nodes and
degree distribution $\mathcal{P}_k$ (fraction of nodes with $k$ links), and it
is allowed to interact with its neighbors in the network. 
A given agent $i$ ($i=1,..,N$) can be in one of two possible states
(playing strategy) $\theta_i=0$ or $\theta_i=1$, corresponding to the
strategies of the mediocre and perfect players, respectively.  In the
initial configuration each agent takes the state $0$ or $1$ with equal
probability $1/2$.  We denote by $\sigma_i$ the fraction of neighbors of agent $i$ that are in state
$\theta=1$.  The expected payoffs of agent $i$ depend on $\sigma_i$, its
state $\theta_i$, and the game (duel or truel) that it plays, and they are given by  
\begin{subequations}
\begin{eqnarray}
P^i_0 &=&  \frac{1}{2} - \frac{1}{4} \sigma_i ~~~ \mbox{if $\theta_i=0$},   \\
P^i_1 &=& \frac{3}{4} - \frac{1}{4} \sigma_i  ~~~ \mbox{if $\theta_i=1$},
\end{eqnarray}
\label{P0-P1-duel}
\end{subequations}
when $i$ plays the duel game, and by
\begin{subequations}
\begin{eqnarray}
\tilde{P}^i_0 &=& \frac{1}{2} \sigma_i^2 + \frac{17}{24} \sigma_i(1-\sigma_i) +
  \frac{1}{3} (1-\sigma_i)^2 ~~~ \mbox{if $\theta_i=0$}, \\
\tilde{P}^i_1 &=& \frac{1}{3} \sigma_i^2 + \frac{1}{2} \sigma_i(1-\sigma_i) +
  \frac{7}{24} (1-\sigma_i)^2 ~~~ \mbox{if $\theta_i=1$},
\end{eqnarray}
\label{P0-P1-truel}
\end{subequations}
when $i$ plays the truel game.  

This system of agents is endowed with the following dynamics.  In a single time step $\Delta t=1/N$, an agent $i$ with state $\theta_i$ is
chosen at random. Then, with probability $p$ agent $i$ plays the duel
game: one neighboring agent $j$ of $i$ is randomly chosen, and $i$ adopts the
state $\theta_j$ of $j$ ($\theta_i \to \theta_i = \theta_j$) with a 
probability equal to $j$'payoff $P_{\theta_j}$. With the complementary probability $1-p$
agent $i$ plays the truel game: two random neighbors $j$ and $k$ of $i$
are chosen, and $i$ tries to adopt either the state $\theta_j$ of $j$
($\theta_i \to \theta_i = \theta_j$) with probability $\frac{1}{2}
\tilde{P}_{\theta_j}$ or the state $\theta_k$ of $k$ ($\theta_i \to
\theta_i = \theta_k$) with probability $\frac{1}{2}
\tilde{P}_{\theta_k}$.  In case that a node with only one neighbor
(degree $k=1$) is chosen to play a truel game, nothing happens.

The system evolves under this dynamics until it reaches a stationary
state, where the fraction of perfect agents remains in a stationary
value $\sigma^{\mbox{\tiny stat}}$.  Note from Eqs.~(\ref{P0-P1-duel}) and (\ref{P0-P1-truel}) that $P^i_1 > P^i_0$ and $\tilde{P}^i_0 > \tilde{P}^i_1$ for all $0 \le \sigma_i \le 1$.
Therefore, in the long run we expect a consensus of state--$1$ agents
(all agents in state $\theta = 1$, $\sigma^{\mbox{\tiny stat}}=1$) when all agents play the duel game, and a state--$0$ consensus when all agents play the truel game ($\sigma^{\mbox{\tiny stat}}=0$).  However, a stationary coexistence of both types of agents might be
possible when agents are allowed to play duel and truel with
probabilities $p$ and $1-p$, respectively.  We are interested in exploring the behavior of
$\sigma^{\mbox{\tiny stat}}$ with $p$ for networks of different degree
distributions, and how $\sigma^{\mbox{\tiny stat}}$ is compared to the value
$\sigma^*$ [Eq.~(\ref{sigma*})] obtained at the mixed equilibrium of strategies predicted in section $IV$.  In the next two sections we study the system on a complete graph
(all-to-all interactions) and on random networks with various degree
distributions, respectively.

\subsection{Complete graph}
\label{CG}

\begin{figure}[t]
  \includegraphics[width=\columnwidth]{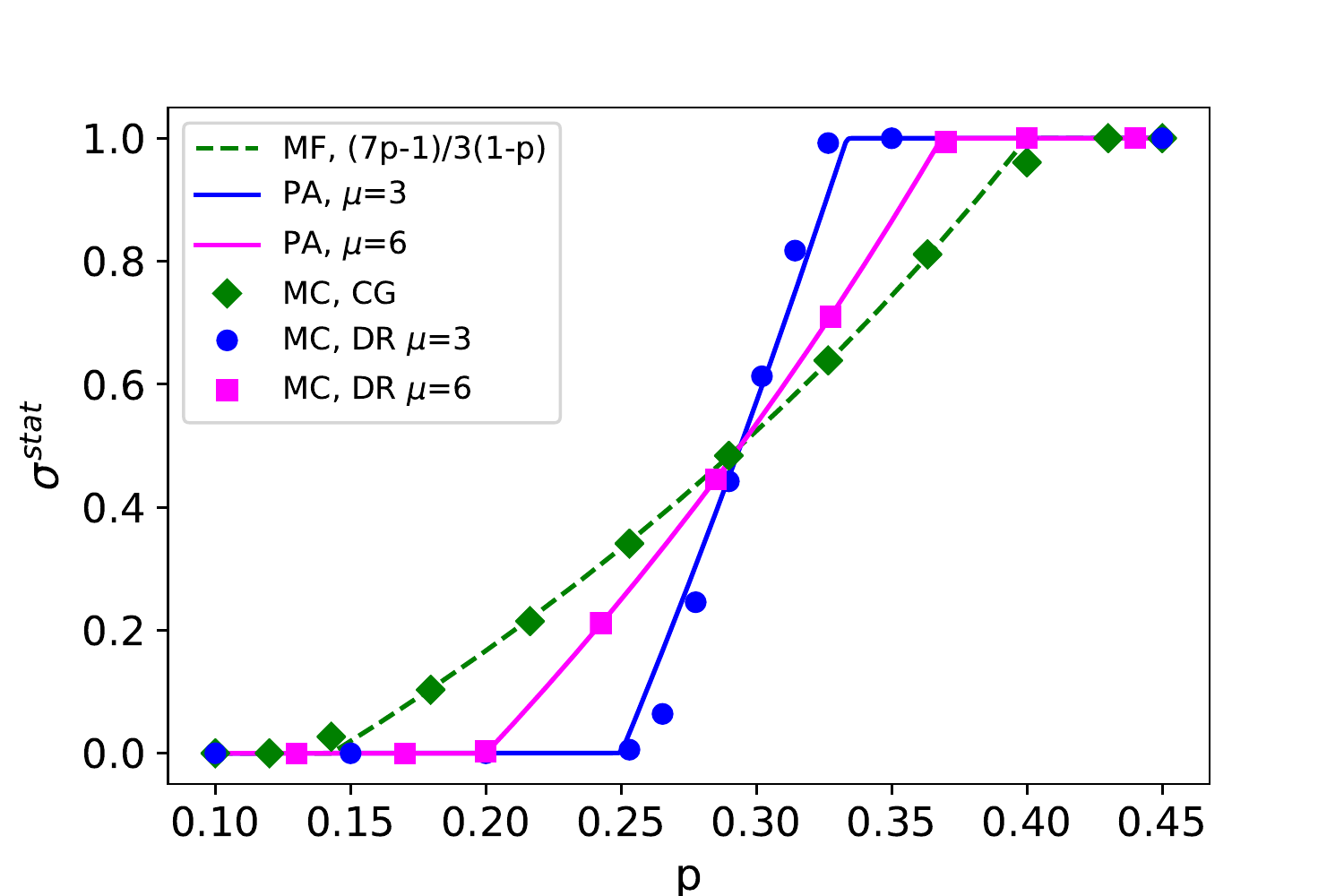}
  \caption{Stationary fraction of perfect agents $\sigma^{\mbox{\tiny stat}}$ vs duel probability $p$, for the values of the mean degree $\mu$ indicated in the legend.  The dashed line corresponds to the stable solution $\sigma_{\mbox{\tiny CG}}$ on a CG [Eq.~(\ref{s-CG}], while solid lines represent the solution from the PA Eqs.~(\ref{dsdt-nets}), (\ref{drhodt}), (\ref{P-duel-truel}) and (\ref{s01-10}).  Symbols correspond to the average value of $\sigma$ at the stationary state obtained from MC simulations on a CG  of size $N=10^3$ (diamonds), and DRRGs of size $N=10^4$ and degrees $\mu=6$ (squares) and $\mu=3$ (circles).}
  \label{s-p-CG-DR}
\end{figure}

In order to gain an insight into the evolution of the system we
investigate in this section the case where each agent can interact
with anyone else, which corresponds to the model on a complete graph
(CG) or mean-field (MF) ($\mathcal{P}_k = \delta_{k,N-1}$).  We define by $\sigma$ the fraction of agents in state $1$ (perfect
players), thus the fraction of state--$0$ agents (mediocre players) is
$1-\sigma$.  In the $N \to \infty$ limit, the time evolution of $\sigma$ is given by the following rate equation:
\begin{eqnarray}
\frac{d\sigma}{dt} = \left[ p (P_1-P_0) + (1-p) (\tilde{P}_1 - \tilde{P}_0)
  \right] \sigma(1-\sigma),
\label{dsdt}
\end{eqnarray}
which becomes
\begin{eqnarray}
  \frac{d\sigma}{dt} = \left[ \frac{1}{4} p - \left( \frac{1}{24} +
  \frac{s}{8} \right) (1-p) \right] \sigma(1-\sigma)
\label{dsdt-1}
\end{eqnarray}
after replacing the expressions for the expected payoffs from
Eqs.~(\ref{P0-P1-duel}) and (\ref{P0-P1-truel}), and taking $\sigma_i=\sigma$,
given that on a CG the fraction of state--$1$ neighbors of a given agent
matches the fraction of state--$1$ agents of the entire population when $N \gg 1$.  The first term in Eq.~(\ref{dsdt}) corresponds to a duel event,
which happens with probability $p$.  The gain term $P_1 \, \sigma(1-\sigma)$
represents the transition $01 \to 11$, where two random agents $i$ and 
$j$ with states $\theta_i=0$ and $\theta_j=1$, respectively, are chosen
with probability $(1-\sigma) \sigma$, and then $i$ copies $j$'s state with
probability $P_1$, leading to a positive change of $1/N$ in $\sigma$.  The loss term
$P_0 \, \sigma(1-\sigma)$ corresponds to the transition $10 \to 00$, where an agent $i$ with state $\theta_i=1$ copies the state of another agent $j$ with state $\theta_j=0$, leading to a negative change of
$1/N$ in $\sigma$.  Analogously, the second term in Eq.~(\ref{dsdt})
corresponds to a truel event (probability $1-p$), where the three
possible transitions that make up the gain term $\tilde{P}_1\,\sigma(1-\sigma)$ are
\begin{subequations}
\begin{eqnarray}
  011 &\to& 111 ~~~ \mbox{with prob. $(1-\sigma) \sigma^2 \, \tilde{P}_1$}, \\
  001 &\to& 101 ~~~ \mbox{with prob. $(1-\sigma)^2 \sigma \, \frac{1}{2} \tilde{P}_1$}, \\
  010 &\to& 110 ~~~ \mbox{with prob. $(1-\sigma)^2 \sigma \, \frac{1}{2} \tilde{P}_1$},   
\end{eqnarray}
\end{subequations}
while the transitions that lead to the loss term $\tilde{P}_0 \, \sigma(1-\sigma)$ are 
\begin{subequations}
\begin{eqnarray}
  100 &\to& 000 ~~~ \mbox{with prob. $\sigma (1-\sigma)^2 \, \tilde{P}_0$}, \\
  110 &\to& 010 ~~~ \mbox{with prob. $\sigma^2 (1-\sigma) \, \frac{1}{2} \tilde{P}_0$}, \\
  101 &\to& 001 ~~~ \mbox{with prob. $\sigma^2 (1-\sigma) \, \frac{1}{2} \tilde{P}_0$}.   
\end{eqnarray}
\end{subequations}
We note that Eq.~(\ref{dsdt-1}) agrees with that of the replicator
Eq.~(\ref{replicatoreq}) derived in section $IV$, which has three fixed points.  The
fixed points $\sigma=0$ and $\sigma=1$ correspond to the consensus of
mediocre and perfect players, respectively, while the third fixed
point $\sigma^*=(7p-1)/[3(1-p)]$ corresponds to a mixed state where state--$1$ and state--$0$ agents coexist with fractions $\sigma^*$ and $1-\sigma^*$, respectively.  An expansion of Eq.~(\ref{dsdt-1}) around $\sigma=0$ to first order in $\epsilon=\sigma \ll 1$ 
leads to $\frac{d\epsilon}{dt} = \frac{(7p-1)}{24} \epsilon$ and,
therefore, $\sigma=0$ is stable for $p<1/7$.  A similar stability
analysis around $\sigma=1$ leads to $\frac{d\epsilon}{dt} =
\frac{(2-5p)}{12} \epsilon$, where $\epsilon = 1-\sigma \ll 1$, and thus
$\sigma=1$ is stable for $p>2/5$.  Then, the stable fixed points on a CG are given by 
\begin{eqnarray} 
\sigma_{\mbox{\tiny CG}} = 
\left\{ \begin{array}{ccc}
    \ 0 & \mbox{for} & p \leq \frac{1}{7}, \\
    \\
    \ \frac{7p-1}{3(1-p)} & \mbox{for} &\frac{1}{7}<p<\frac{2}{5}, \\
    \\
    \ 1 & \mbox{for} & p \ge \frac{2}{5}.
\end{array}\right.
\label{s-CG}
\end{eqnarray}
In Fig.~\ref{s-p-CG-DR} we plot by a dashed line the stable fixed points $\sigma_{\mbox{\tiny CG}}$ of Eq.~(\ref{dsdt-1}) as a function of $p$, which corresponds to the dynamics on a CG.  This behavior is the same as that obtained from the theory of Nash equilibrium analyzed in section $IV$.  Therefore, we conclude that the Nash equilibrium corresponds to the stable states of the agent-based model on a CG or MF scenario.   

These results show that a stable coexistence of perfect and mediocre players is obtained for values of $p$ in the interval $p_0 < p < p_1$, with $p_0=1/7$ and $p_1=2/5$, while for $p<p_0$ ($p>p_1$) mediocre (perfects) players dominate.  In order to explore how the coexistence phase is affected by the topology of interactions between agents, we study in the next section the duel-truel dynamics on complex networks.

\subsection{Complex networks}
\label{networks}

In this section we derive and analyze equations that describe the time evolution
of the system in complex networks with a general degree distribution
$\mathcal{P}_k$, i.e., the fraction of nodes with $k$ links, subject to
the normalization $\sum_{k=1}^{N-1} \mathcal{P}_k=1$.
For that, we use the \emph{homogeneous pair approximation} (PA)
approach developed in \cite{Vazquez-2008-1,Vazquez-2008-2,Vazquez-2010,Demirel-2014},
which is a mean-field approach that takes
into account state correlations between first nearest neighbors in the
network, and neglects correlations to second and higher--order nearest
neighbors.  This approximation should work well in uncorrelated or
random networks like degree-regular random graphs (DRRG) or Erd\"os-Renyi (ER)
networks, where the neighbors of each node are chosen at random, and
thus degree correlations are negligible.  In Appendix~\ref{s-rho} we derive
a rate equation for the fraction of perfect agents $\sigma$ that depends on
$\sigma$ and the fraction of \emph{active links} $\rho$, i.e., links that
connect mediocre and perfect agents ($0$--$1$ or $1$--$0$ links) and, therefore, are subject to change.  The
density $\rho$ allows to capture correlations between the states of
neighboring nodes in the network, and together with $\sigma$ form the
following system of two rate equations that describe the dynamics of
the model on a complex network (see Appendix~\ref{s-rho}): 
\begin{eqnarray}
  \label{dsdt-nets}
  \frac{d\sigma}{dt} &=& \frac{\rho}{2} \left[ p (P_{1|0}-P_{0|1}) +
                    (1-p)(\tilde{P}_{1|0} - \tilde{P}_{0|1})
                    (1-\mathcal{P}_1) \right], \nonumber \\
\end{eqnarray}
\begin{eqnarray}
  \label{drhodt}
    \frac{d\rho}{dt} &=& \frac{\rho}{\mu} \Bigg\{ \left[ p P_{0|1} +
                         (1-p) \tilde{P}_{0|1} \right] \left[ (\mu-1)
                         \left(1-\frac{\rho}{\sigma} \right) - 1 \right]
                         \nonumber \\
  &+& \left[ p P_{1|0} + (1-p) \tilde{P}_{1|0}
                         \right] \left[ (\mu-1) \left(
                         1-\frac{\rho}{1-\sigma} \right) - 1 \right]
                         \nonumber \\
    &+& (1-p) \left( \tilde{P}_{0|1} + \tilde{P}_{1|0} \right) \mathcal{P}_1 \Bigg\},
\end{eqnarray}
where $\mu \equiv \langle k \rangle = \sum_{k=1}^{N-1} k \, \mathcal{P}_k$
is the mean degree of the network.  The terms that involve $\mathcal{P}_1$ correspond to the case where an agent of degree $k=1$ fails to play a truel game.  Here we denote by $P_{0|1}$ and
$\tilde{P}_{0|1}$ the expected payoffs of a state--$0$ agent that has at
least one neighbor in state $1$, when it plays a duel and a truel,
respectively, and analogously for $P_{1|0}$ and $\tilde{P}_{1|0}$.  The expected payoffs are given by  Eqs.~(\ref{P0-P1-duel}) and (\ref{P0-P1-truel}):
\begin{subequations}
\begin{eqnarray}
    P_{0|1} &=& \frac{1}{2} - \frac{1}{4} \sigma_{0|1},  \\
    P_{1|0} &=& \frac{3}{4} - \frac{1}{4} \sigma_{1|0},  \\
    \tilde{P}_{0|1} &=& \frac{1}{2} \sigma_{0|1}^2 + \frac{17}{24} \sigma_{0|1}(1-\sigma_{0|1}) + \frac{1}{3} (1-\sigma_{0|1})^2,  \\
    \tilde{P}_{1|0} &=& \frac{1}{3} \sigma_{1|0}^2 + \frac{1}{2}
                        \sigma_{1|0}(1-\sigma_{1|0}) + \frac{7}{24}
                        (1-\sigma_{1|0})^2, 
\end{eqnarray}
\label{P-duel-truel}
\end{subequations}
where $\sigma_{0|1}$ ($\sigma_{1|0}$) is the fraction of state--$1$ neighbors of
a state--$0$ (state--$1$) node that has at least one neighbor in the
opposite state $1$ ($0$), which can be estimated as (see Appendix~\ref{Eq-s})
\begin{subequations}
\begin{eqnarray}
    \sigma_{0|1} &\simeq& \frac{1}{\mu} \left[ 1 + \frac{(\mu-1)\rho}{2(1-\sigma)}
  \right] ~~~ \mbox{and} \\
  \sigma_{1|0} &\simeq& \frac{(\mu-1)}{\mu} \left( 1 - \frac{\rho}{2\sigma} \right).
\end{eqnarray}
\label{s01-10}
\end{subequations}

Now we can use these equations to study how the mean degree $\mu$ of the network and $\mathcal{P}_1$ affect the behavior of $\sigma$ with $p$ at the stationary state.  For that, we integrated numerically the set of Eqs.~(\ref{dsdt-nets}) and (\ref{drhodt}) together with Eqs.~(\ref{P-duel-truel}) and (\ref{s01-10}), and plot the stationary value $\sigma^{\mbox{\tiny stat}}$ vs p for different values of $\mu$ and $\mathcal{P}_1$, as it is shown by lines in Fig.~\ref{s-p-CG-DR}.  In this figure we see the results for a degree distribution $\mathcal{P}_k=~\delta_{k,\mu}$, with $\mathcal{P}_1=0$ and $\mu=6$ and $3$ (solid lines), which corresponds to DRRGs  where each node is connected to other $\mu$ random nodes.  We have also checked that the solution for the case $\mathcal{P}_k=\delta_{k,N-1}$ matches that of the CG case obtained from Eq.~(\ref{dsdt-1}), and depicted by a dashed line.  We observe that, as $\mu$ decreases the transition values $p_0$ and $p_1$ that define the coexistence phase become closer and, therefore, the interval of $p$ for which there is a coexistence of perfect and mediocre players is reduced. This suggests that the coexistence phase tends to shrink and might eventually vanish as the network becomes more sparse by reducing $\mu$.

Although these results were obtained from the PA equations and for DRRGs, we shall see in the next section that a similar behavior is observed for networks with a broader degree distribution.

\subsection{Monte Carlo simulations}
\label{simulations}

We have performed extensive numerical simulations on various networks and checked the analytical results obtained in sections \ref{CG} and \ref{networks}.  We started by running the Monte Carlo (MC) dynamics described in section~\ref{model} on a CG of $N=10^3$ nodes, and DRRGs of $N=10^4$ nodes each and mean degrees $\mu=3$ and $\mu=6$.  We ran the dynamics until the fraction of perfect agents $\sigma$ reached a stationary value, and calculated the average value of $\sigma$ at the stationary state over many independent realizations.  Results are shown by symbols in Fig.~\ref{s-p-CG-DR}.  

\begin{figure}[t]
  \includegraphics[width=\columnwidth]{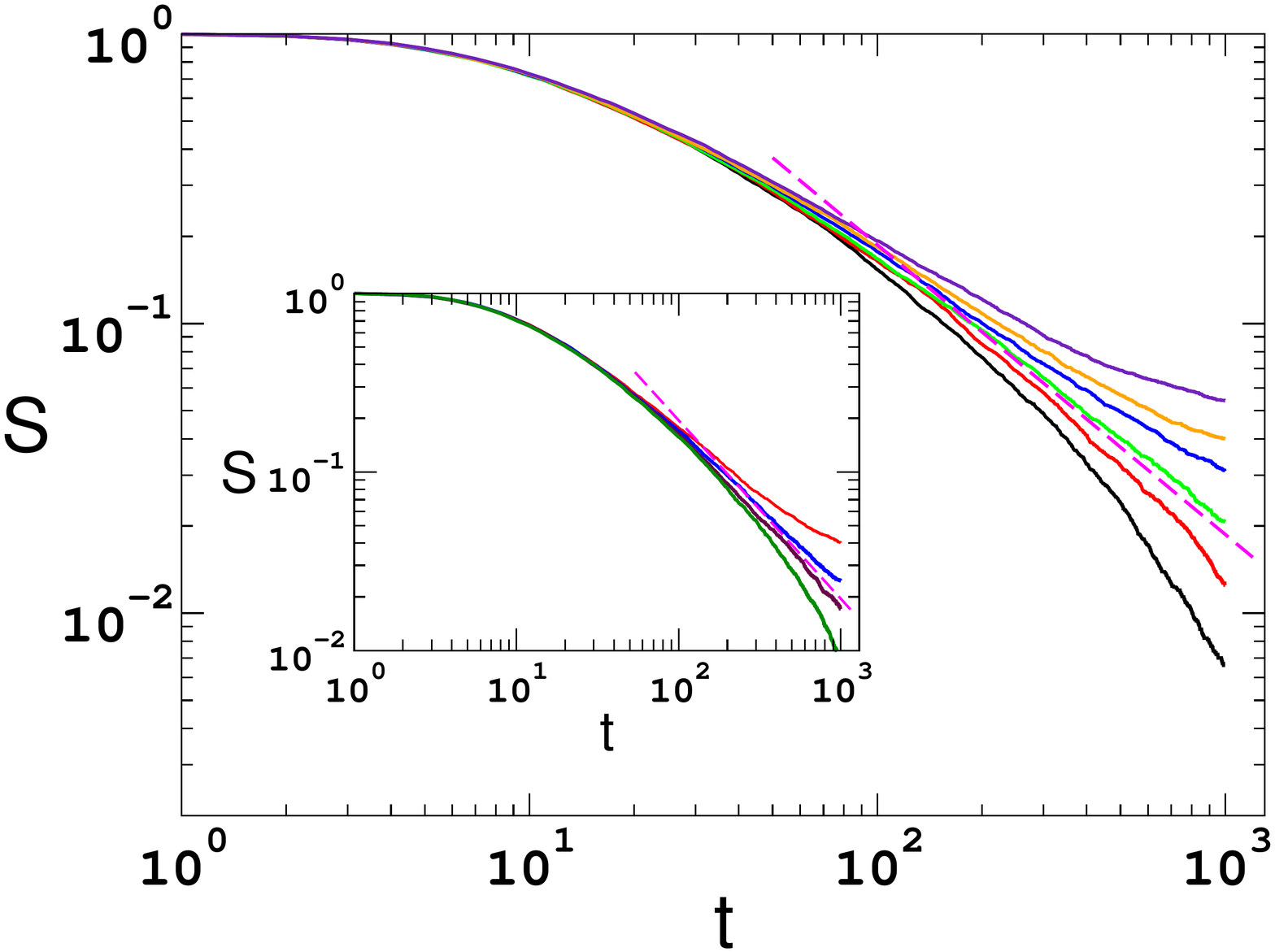}
  \caption{Survival probability $S$ on a DRRG of $N=10^4$ nodes, mean degree $\mu=3$ and various values of $p$.  Main panel: evolution of $S$ after a small perturbation of the $0$--consensus absorbing state.  Curves correspond to $p=0.265, 0.260, 0.255, 0.250, 0.245$ and $0.240$ (from top to bottom).   Inset: evolution of $S$ after a small perturbation of the $1$--consensus absorbing state.  Curves correspond to $p=0.325, 0.330, 0.335$ and $0.340$ (from top to bottom).  At the transition points $p_0 \simeq 0.250$ and $p_1 \simeq 0.335$, $S$ decays as $S \sim t^{-1}$ (dashed lines).}
  \label{surv}
\end{figure}

We observe a good agreement with the analytical results provided by the MF approximation $\sigma_{\mbox{\tiny CG}}$ on a CG (dashed line), and by the PA from Eqs.~(\ref{dsdt-nets}) and (\ref{drhodt}) (solid lines).  However, we see that close to the transition points $p_0$ and $p_1$ the simulation points slightly deviate from the analytical prediction.  To check if this is due to finite-size effects, we have performed spreading experiments that allowed to calculate the transition points with high accuracy.  These experiments consisted on making a small perturbation of the two absorbing states corresponding to $0$--consensus and $1$--consensus, and see how this perturbation evolves.  For instance, for the $\sigma=0$ absorbing state we assigned initially the state $\sigma=0$ to all agents except for a small seed of four random neighboring agents in the network that were assigned the state $\sigma=1$, and let the system evolve.  Then, for values of $p$ below the transition point $p_0$ the seed quickly vanishes and thus all realizations quickly reach the absorbing state, while for $p>p_0$ there are some realizations that survive, where the seed spreads over a fraction of the population.  An analogous behavior is obtained around $p_1$ by perturbing the $\sigma=1$ absorbing state.  To quantify this we ran the dynamics with these initial conditions and measured the survival probability $S(t)$ of a single run, calculated as the fraction of realizations that did not reach the absorbing state up to time $t$.  Then, the transition point is estimated as the value of $p$ for which $\sigma$ decays as a power law.  In Fig.~\ref{surv} we plot the time evolution of $S(t)$ for a DRRG of size $N=10^4$ and $\mu=3$.  We see that $\sigma$ decays as a power law with exponent $-1$ at the transition points $p_0 \simeq 0.250$ (main panel) and $p_1 \simeq 0.335$ (inset).  These values are very similar to those obtained from the PA, $p_0 \simeq 0.248$ and $p_1 \simeq 0.338$, which confirms that the discrepancies with the numerical simulations in Fig.~\ref{s-p-CG-DR} are due to finite-size effects.

\begin{figure}[t]
  \includegraphics[width=\columnwidth]{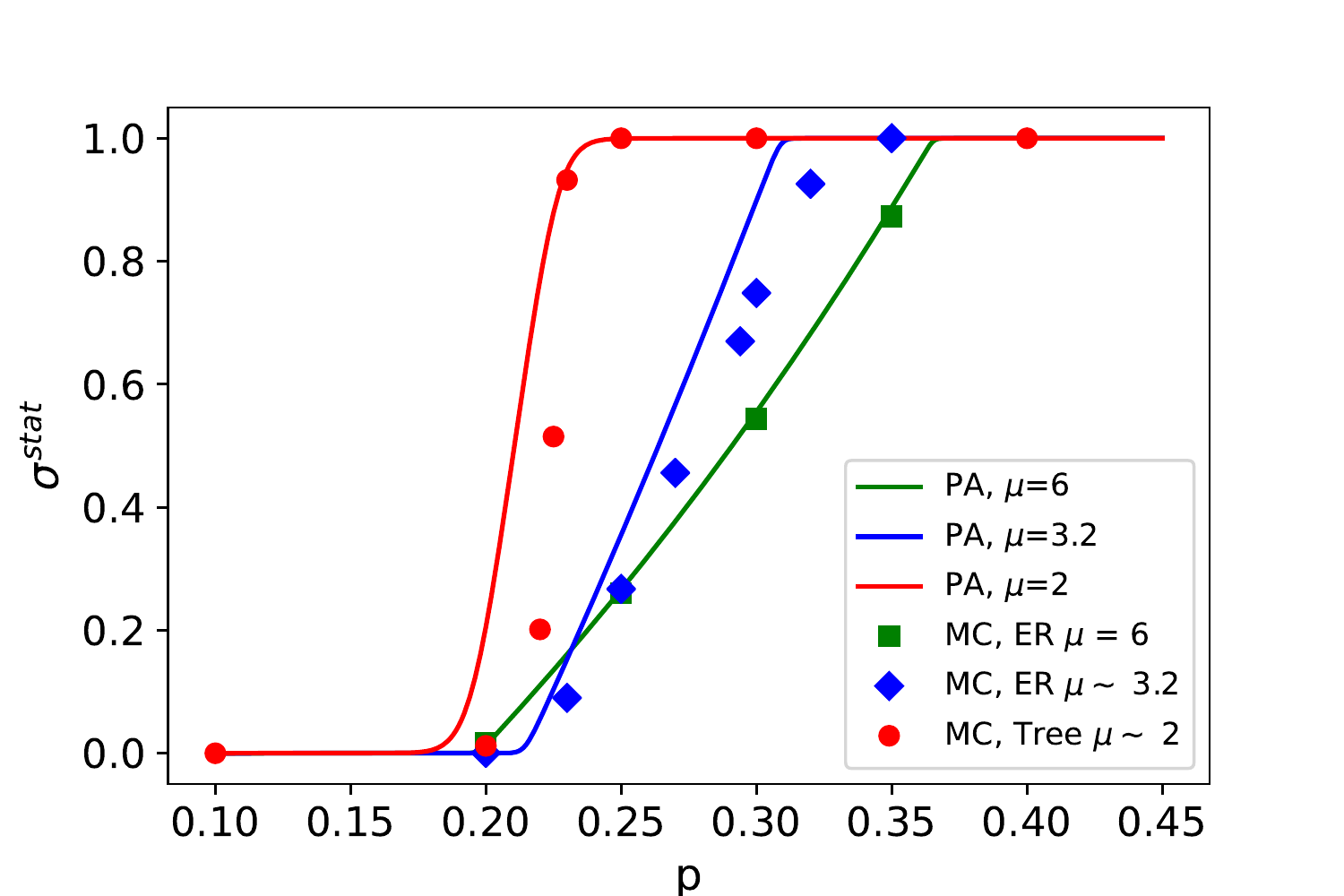}
  \caption{$\sigma^{\mbox{\tiny stat}}$ vs $p$ for the values of $\mu$ and $\mathcal{P}_1$ indicated in the legend.  Solid lines are the solutions from the PA, while symbols correspond to the average value of $\sigma$ at the stationary state obtained from MC simulations on ER networks with $N=10^4$ nodes and mean degrees $\mu=6$ (squares) and $\mu \simeq 3.2$ (diamonds), and on a tree network with $N=10^4$ nodes and $\mu \simeq 2$ (circles).}
  \label{s-p-ER}
\end{figure}

We then ran simulations on Erd\"os-Renyi (ER) networks to explore the impact of the broadness of the degree distribution on the results shown above.  We were particularly interested in the case of low mean degrees, where the behavior tends to be quite different from that of MF, and where analytical approximations usually fail.   As ER networks with low mean degree may get disconnected, we considered the largest connected component of the network as the new system, and used the values of its resulting mean degree $\mu$ and $\mathcal{P}_1$ in the PA equations.  We also ran simulations on networks with a tree-like structure, which have a mean degree close to $2$.  In Fig.~\ref{s-p-ER} we show by symbols the results of MC simulations for ER networks of $N=10^4$ nodes each and mean degrees $\mu=6$ and $\mu=3.2$, and for tree-like networks of $N=10^4$ nodes and $\mu \simeq 2$.  We also show by solid lines the solutions from the PA Eqs.~(\ref{dsdt-nets}) and (\ref{drhodt}), with the values of $\mu$ and $\mathcal{P}_1$ that correspond to each network, as shown in the legend.  We see that the agreement with simulations is good for the ER network with $\mu=6$, but discrepancies arise for the ER with $\mu=3.2$ and the tree network, which confirms that the PA start to fail for very low values of $\mu$. 

\begin{figure}[t]
  \includegraphics[width=\columnwidth]{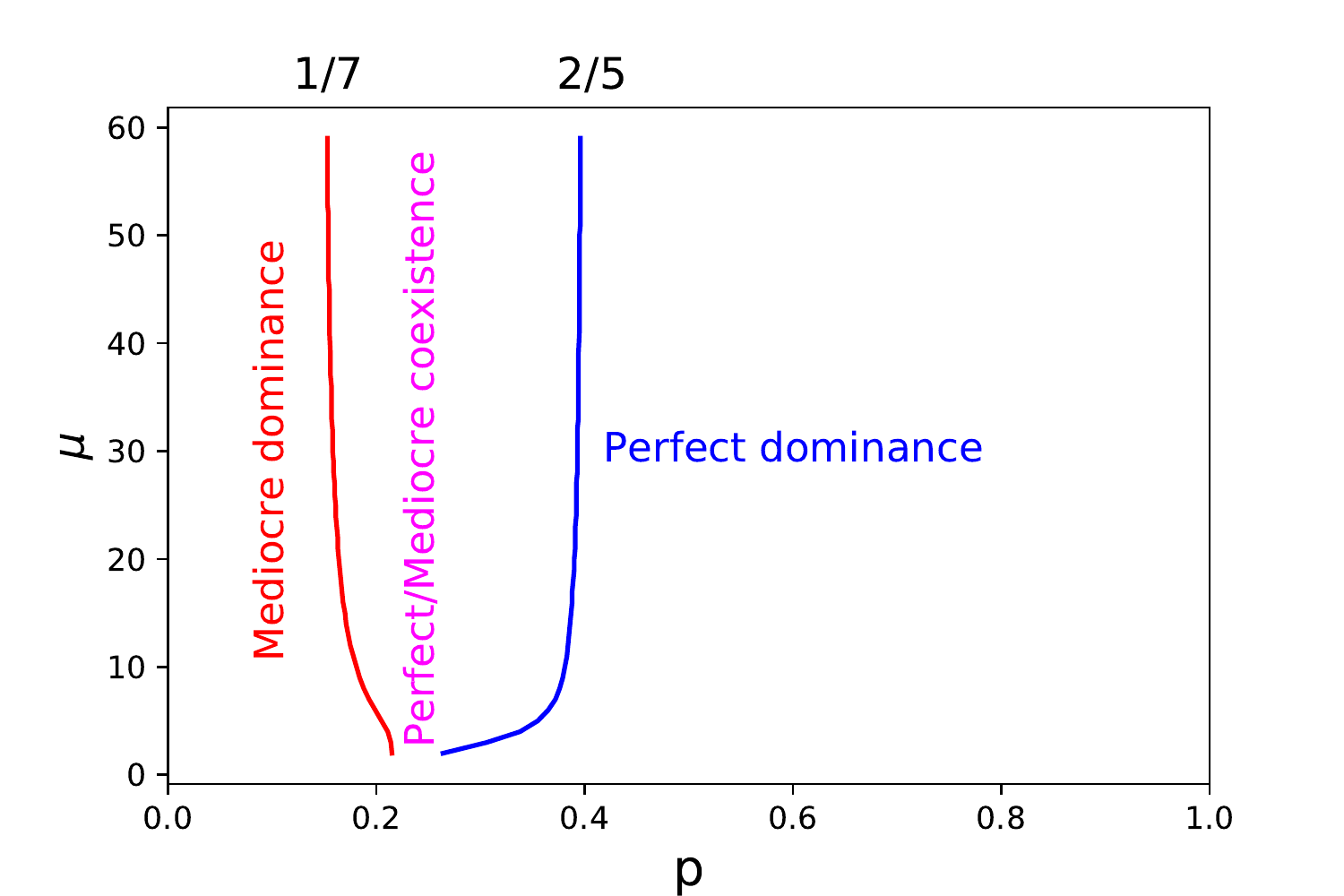}
  \caption{Phase diagram on the $p$--$\mu$ space showing the transition lines between the coexistence and dominance phases, obtained from the PA Eqs.~(\ref{dsdt-nets}) and (\ref{drhodt}) for an ER network.}
  \label{phase-diag}
\end{figure}

We can also see in Fig.~\ref{s-p-ER} that the coexistence phase becomes smaller when $\mu$ decreases, as we have seen already for DRRGs (Fig.~\ref{s-p-CG-DR}), but it does not seem to vanish completely.  For instance, the coexistence phase for the tree network ($\mu \simeq 2$) lays in the small interval $0.20 \lesssim p \lesssim 0.25$.  To explore this in more detail, we integrated Eqs.~(\ref{dsdt-nets}) and (\ref{drhodt}) for several values of $\mu \ge 2$ and $\mathcal{P}_1$ corresponding to an ER network, and obtained the transition lines $p_0(\mu)$ and $p_1(\mu)$ that separate the coexistence and dominance phases.  This is shown in the phase diagram of Fig.~\ref{phase-diag}, where we see that the coexistence phase increases with $\mu$ and approaches the region defined by the interval $[1/7, 2/5]$ that corresponds to the CG.

\section{Summary and conclusions}
\label{conclusions}

In this paper we have extended the definition of evolutionary stable strategy to symmetric games with random number of players.  We derived the conditions under which a given strategy is an ESS, and applied this concept to the study of a population of players that can use one of two possible strategies, perfect and mediocre, and play a duel with probability $p$ or a truel with probability $1-p$.  We showed that the duel-truel game has a mixed equilibrium that is an ESS for values of $p$ in an interval, in which perfect and mediocre players coexist.  

We also introduced and studied an agent-based model on complex networks with a microscopic dynamics that leads to an evolution that is well described in mean-field by the replicator equation.  We showed that the stable solutions of this equation correspond to those of the ESS equilibrium. Moreover, we developed a pair approximation approach to study the dynamics of the model on complex networks with a general degree distribution, which showed that the coexistence phase predicted by the theory is also present when agents interact in complex topologies.  We applied this approach to random networks and found that the interval of $p$ for which the stable coexistence of perfect and mediocre players is observed only depends on the mean degree of the network $\mu$, but not on higher moments.  We found that the coexistence phase shrinks as $\mu$ decreases, but it does not seem to vanish completely even for small values of $\mu$.  As a consequence, a given unstable mix of the two types of players for some value of $p$, can turn into stable when the mean number of neighbors of a player is increased beyond a threshold.  This result implies that the network of interactions affects the stability of the system by inducing a stable coexistence when its connectivity increases.     

In order to check these findings we performed Monte Carlo simulations of the model in degree regular random graphs of low mean degrees, and we could verified that the analytical results were in good agreement with simulations on these networks.  The accuracy of the transition points between the coexistence and dominance phases predicted by the pair approximation was also tested by means of spreading experiments.  We also carried out simulations on Erd\"os-Renyi and tree-like networks to investigate the behavior of the model for broad degree distributions and $\mu$ close to $2$.  We found that for low degrees the analytical approximation gives results that deviate from simulations.  However, we could verified numerically that, although very small, the coexistence phase is still present in sparse networks.     

It would be worthwhile to perform a more in-depth study of the duel-truel game in complex networks with very broad degree distributions, like scale-free networks, and investigate if the observed phenomenology is affected by the width of the distribution.  It might also be interesting to extend the theoretical approximations for networks that have degree correlations.  Finally, as a future work we could apply the conditions obtained for ESS to games with interactions higher than three and different strategies. 


\section*{ACKNOWLEDGMENTS}

This work was partially supported by ANPCyT under grants PICT 201--0215 and 
PICT 2016--1022, and by Universidad de Buenos Aires under grant UBACYT 2018 20020170100445 BA. 
\\

\section*{DATA AVAILABILITY}
The data that support the findings of this study are available
within the article.

\section*{DECLARATIONS}
 
{\bf Conflict of interest.} The authors  have no conflict of interest to disclose.

\appendix

\begin{widetext}

\section{Derivation of the rate equations for $\sigma$ and $\rho$}
\label{s-rho}

In this section we derive rate equations for the evolution of the fraction of perfect agents $\sigma$ and the fraction of active links $\rho$ on random networks with a general degree distribution $\mathcal{P}_k$. 

\subsection{Equation for $\sigma$}
\label{Eq-s}

In Fig.~\ref{update}(a) we show an schematic illustration of the transition $01 \to 11$ in a single update event, which leads to a change of $1/N$ in $\sigma$.  This transition involves the following processes and associated probabilities. A node $i$ with state $\theta_i=0$ and degree $k$ is
chosen at random with probability $(1-\sigma) \mathcal{P}_k$.  Then, with
probability $n_{0,k}^1/k$ a neighbor $j$ with state $\theta_j=1$ is chosen at random, where $n_{0,k}^1$ ($0 \le n_{0,k}^1 \le k$) is the number of neighbors of $i$
in the opposite state $\theta=1$, i.e., the number of \emph{active links}
connected to node $i$.  We recall that an active link is a link
that connects nodes with different states ($0$--$1$ or $1$--$0$ links).
Finally, $i$ copies $j$'s state with a probability equal to the mean payoff of $j$, which we
denote by $P_{1|0}$ or $\tilde{P}_{1|0}$, when $i$ plays a duel or
truel, respectively.  Here we use the subindex $1|0$ to denote that
$j$ has state $\theta_j=1$ and at least one neighbor (node $i$) with state
$\theta_i=0$ [see Fig.~\ref{update}(a)].  

\begin{figure}[t]
  \includegraphics[width=10cm]{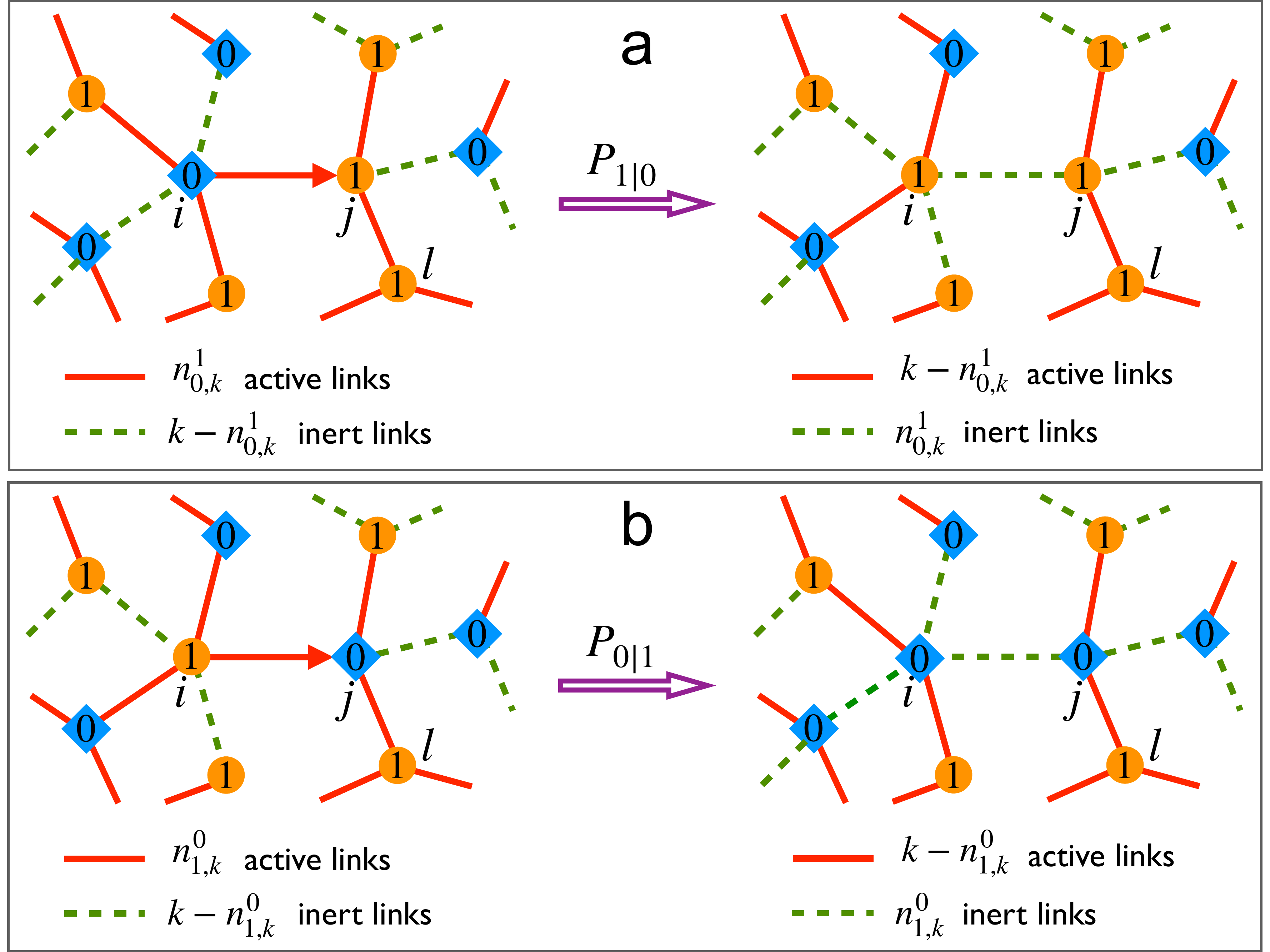}
  \caption{a) Update event in which a node $i$ with state $\theta_i=0$ (diamond) adopts the state $\theta_j=1$ of its neighboring node $j$ (circle) with probability $P_{1|0}$ ($j$'s payoff).  The change in $\sigma$ and $\rho$ are $1/N$ and $\frac{2(k-2n_{0,k}^1)}{\mu N}$, respectively.  b) Update event in which a node $i$ with state $\theta_i=1$ (circle) adopts the state $\theta_j=0$ of its neighboring node $j$ (diamond) with probability $P_{0|1}$ ($j$'s payoff).  The change in $\sigma$ and $\rho$ are $-1/N$ and $\frac{2(k-2n_{1,k}^0)}{\mu N}$, respectively. }
  \label{update}
\end{figure}

Analogously, with probability $\sigma \, \mathcal{P}_k \,
\frac{n_{1,k}^0}{k} \, P_{0|1}$ a randomly chosen node $i$ with
state $\theta_i=1$ and degree $k$ copies a random neighbor $j$ with
state $\theta_j=0$ when it plays a duel [see Fig.~\ref{update}(b)], where $n_{1,k}^0$ is the number
of state--$0$ neighbors of $i$ and $P_{0|1}$ is the expected payoff of
$j$, which has at least a neighbor with state $1$ (node $i$).  This
corresponds to a $10 \to 00$ 
transition, leading to a change of $-1/N$ in $\sigma$.  The expression for
the probability of a truel game is the same but replacing $P_{0|1}$ by
$\tilde{P}_{0|1}$.  Adding these four terms corresponding to the two
transitions and the two playing strategies, we can write the mean change of $\sigma$ in a time step $\Delta t=1/N$ as  
\begin{eqnarray}
	\frac{d\sigma}{dt} &=& \frac{(1-\sigma)}{1/N} \Bigg\{ p \sum_{k=1}^{N-1} \mathcal{P}_k \sum_{n_{0,k}^1=0}^k B \left( n_{0,k}^1 \right) \frac{n_{0,k}^1}{k}  \, P_{1|0} + (1-p) \sum_{k=2}^{N-1} \mathcal{P}_k \sum_{n_{0,k}^1=0}^k B \left( n_{0,k}^1 \right) \frac{n_{0,k}^1}{k} \, \tilde{P}_{1|0} \Bigg\}  \frac{1}{N} \nonumber \\ 
	&+& \frac{\sigma}{1/N} \Bigg\{ p \sum_{k=1}^{N-1} \mathcal{P}_k \sum_{n_{1,k}^0=0}^k B \left( n_{1,k}^0 \right) \frac{n_{1,k}^0}{k}  \, P_{0|1} + (1-p) \sum_{k=2}^{N-1} \mathcal{P}_k \sum_{n_{1,k}^0=0}^k B \left( n_{1,k}^0 \right) \frac{n_{1,k}^0}{k} \, \tilde{P}_{0|1} \Bigg\} \left( -\frac{1}{N} \right),
\label{ds-dt}
\end{eqnarray}
where $B\left( n_{0,k}^1 \right)$ is the probability that $n_{0,k}^1$
active links are connected to a node of state $0$ and degree $k$; and
analogously for $B \left( n_{1,k}^0 \right)$.  The first and second
summations in each term run over the degrees $k$ and the number of
active links around a node of degree $k$, respectively.   We note that
the summations over $k$ in the second and fourth terms, which
corresponds to a truel game, start from $k=2$ because a node with only
one neighbor (degree $k=1$) cannot play the truel game.  The mean
payoffs of a neighbor $j$ of $i$ are given by Eqs.~(\ref{P0-P1-duel}) and (\ref{P0-P1-truel}):
\begin{subequations}
\begin{eqnarray}
P_{0|1} &=& \frac{1}{2} - \frac{1}{4} \sigma_{0|1}, \\
\tilde{P}_{0|1} &=& \frac{1}{2} \sigma_{0|1}^2 + \frac{17}{24} \sigma_{0|1}(1-\sigma_{0|1}) +
\frac{1}{3} (1-\sigma_{0|1})^2,
\end{eqnarray}
\label{P01}
\end{subequations}
if $\theta_j=0$ and $\theta_i=1$, and  
\begin{subequations}
\begin{eqnarray}
P_{1|0} &=& \frac{3}{4} - \frac{1}{4} \sigma_{1|0}, \\
\tilde{P}_{1|0} &=& \frac{1}{3} \sigma_{1|0}^2 + \frac{1}{2} \sigma_{1|0}(1-\sigma_{1|0}) +
\frac{7}{24} (1-\sigma_{1|0})^2,
\end{eqnarray}
\label{P10}
\end{subequations}
if $\theta_j=1$ and $\theta_i=0$, where we denote by $\sigma_{0|1}$ and
$\sigma_{1|0}$ the fraction of state--$1$ neighbors of $j$ when it has
state $\theta_j=0$ and $\theta_j=1$, respectively.  These fractions can be estimated as
\begin{equation}
  	\sigma_{0|1} \simeq \sum_{k'=1}^{N-1} \frac{k'
          \mathcal{P}_{k'}}{\mu}   \frac{\langle n_{0,k'}^1
          \rangle}{k'} ~~~ \mbox{and} ~~~ 
	\sigma_{1|0} \simeq \sum_{k'=1}^{N-1} \frac{k' \mathcal{P}_{k'}}{\mu}   \frac{\langle n_{1,k'}^1 \rangle}{k'},
        \label{sj}
\end{equation}   
where $\langle n_{0,k'}^1 \rangle$  ($\langle n_{1,k'}^1 \rangle$)
is the mean number of neighbors of $j$ that are in state $1$ when
$\theta_j=0$ ($\theta_j=1$).  We have also used that $j$ has degree $k'$ with
probability $k' \, \mathcal{P}_{k'}/\mu$, given that in ucorrelated networks the
probability that a random node (node $i$) is connected to a node of degree
$k'$ (node $j$) is proportional to its degree. Then, the mean numbers $\langle
n_{0,k'}^1 \rangle$ and $\langle n_{1,k'}^1 \rangle$ can be approximated as 
\begin{subequations}
  \begin{eqnarray}
        \label{n0k1}
	\langle n_{0,k'}^1 \rangle &\simeq& 1 + (k'-1)P(1|01) ~~~ \mbox{and} \\
        \label{n1k1}    
	\langle n_{1,k'}^1 \rangle &\simeq& (k'-1)P(1|10), 
\end{eqnarray}
\label{n0k}
\end{subequations}
where $P(1|01)$ ($P(1|10)$) is the conditional probability that a neighbor $l \ne
i$ of $j$ has state $\theta_l=1$ given that $\theta_j=0$ ($\theta_j=1$) and
$\theta_i=1$ ($\theta_i=0$) (see Fig.~\ref{update}).  Equations~(\ref{n0k1}) and
(\ref{n1k1}) take into account that $j$ has at least one neighbor (node
$i$) with state $1$ and $0$, respectivley.  Within a PA, $P(1|01)$ and
$P(1|10)$ can be approximated as $P(1|0)$ and $P(1|1)$, respectively,
if we neglect correlations between second nearest neighbors, i.e.,
between $\theta_l$ and $\theta_i$.  Then, if we denote by $\rho$ the
fraction of active links, $P(1|0)$ and $P(1|1)$ can be calculated from the Bayes' equality
\begin{equation}
  \rho = P(0) P(1|0) + P(1) P(0|1),
  \label{rho}
\end{equation}
where $P(0)=(1-\sigma)$ and $P(1)=\sigma$ are the probabilities that a randomly
chosen node is in state $0$ and $1$, respectively.  As the two terms in right hand side of Eq.~(\ref{rho}) are equal, we have that 
\begin{subequations}
\begin{eqnarray}
  P(1|0) &=& \frac{\rho}{2(1-\sigma)}, \\
  P(0|1) &=& \frac{\rho}{2 \sigma},
\end{eqnarray}
\label{P10-01}
\end{subequations}
and thus
\begin{eqnarray}
  P(1|1) = 1 - P(0|1) = 1 - \frac{\rho}{2 \sigma}.
\end{eqnarray}  
Using these expressions for $P(1|0)$ and $P(1|1)$ we obtain from
Eqs.~(\ref{n0k}) that
\begin{subequations}
  \begin{eqnarray}
        \label{n0k1-1}
	\langle n_{0,k'}^1 \rangle &\simeq& 1 + (k'-1)\frac{\rho}{2(1-\sigma)}, ~~~ \mbox{and} \\
        \label{n1k1-1}    
	\langle n_{1,k'}^1 \rangle &\simeq& (k'-1)\left(1-\frac{\rho}{2 \sigma}\right). 
\end{eqnarray}
\label{n0k-1}
\end{subequations}
Plugging these expressions into Eqs.~(\ref{sj}) and performing the
summations we finally obtain
\begin{eqnarray}
  \sigma_{0|1} \simeq \frac{1}{\mu} \left[ 1 + \frac{(\mu-1)\rho}{2(1-\sigma)} \right]
  ~~~ \mbox{and} ~~~
  \sigma_{1|0} \simeq \frac{(\mu-1)}{\mu} \left(1-\frac{\rho}{2 \sigma} \right),  
\label{s01-10-1}
\end{eqnarray}  
which are quoted in Eq.~(\ref{s01-10}) of the main text.

As $\sigma_{0|1}$ and $\sigma_{1|0}$ only depend on $\sigma$, $\rho$ and $\mu$, and
so the payoffs from Eqs.~(\ref{P01}) and (\ref{P10}), we can pull
$P_{1|0}$, $\tilde{P}_{1|0}$, $P_{0|1}$ and $\tilde{P}_{0|1}$ out of
the summations of Eq.~(\ref{ds-dt}).  Then, expressing the second and fourth summations over $k$ as $\sum_{k=2}^{N-1} \mathcal{P}_k=
\sum_{k=1}^{N-1} \mathcal{P}_k - \mathcal{P}_1$, we can write 
\begin{eqnarray}
	\frac{d\sigma}{dt} &=& (1-\sigma) \Bigg\{ \left[ p P_{1|0} + (1-p)
                          \tilde{P}_{1|0}  \right] \sum_{k=1}^{N-1}
                          \mathcal{P}_k \sum_{n_{0,k}^1=0}^k B \left(
                          n_{0,k}^1 \right) \frac{n_{0,k}^1}{k}  -
                          (1-p) \tilde{P}_{1|0} \mathcal{P}_1
                          \sum_{n_{0,1}^1=0}^1 B \left( n_{0,k}^1
                          \right) n_{0,1}^1 \Bigg\}  \nonumber \\ 
  &-& \sigma \Bigg\{ \left[ p P_{0|1} + (1-p)
                          \tilde{P}_{0|1}  \right] \sum_{k=1}^{N-1}
                          \mathcal{P}_k \sum_{n_{1,k}^0=0}^k B \left(
                          n_{1,k}^0 \right) \frac{n_{1,k}^0}{k}  -
                          (1-p) \tilde{P}_{0|1} \mathcal{P}_1
                          \sum_{n_{1,1}^0=0}^1 B \left( n_{1,1}^0
      \right) n_{1,1}^0 \Bigg\} \\
&=& (1-\sigma) \Bigg\{ \left[ p P_{1|0} + (1-p)
                          \tilde{P}_{1|0}  \right] \sum_{k=1}^{N-1}
                          \mathcal{P}_k \frac{\langle n_{0,k}^1 \rangle}{k}  -
                          (1-p) \tilde{P}_{1|0} \mathcal{P}_1
                          \langle n_{0,1}^1 \rangle \Bigg\}  \nonumber \\ 
  &-& \sigma \Bigg\{ \left[ p P_{0|1} + (1-p)
                          \tilde{P}_{0|1}  \right] \sum_{k=1}^{N-1}
                          \mathcal{P}_k  \frac{\langle n_{1,k}^0 \rangle}{k}  -
                          (1-p) \tilde{P}_{0|1} \mathcal{P}_1
                          \langle n_{1,1}^0 \rangle \Bigg\},
      \label{ds-dt-1}
\end{eqnarray}
where $\langle n_{0,k}^1 \rangle$ and $\langle n_{1,k}^0 \rangle$ are
the first moments of $B \left(n_{0,k}^1 \right)$ and $B \left(
  n_{1,k}^0 \right)$, respectively.  These moments can be calculated
by means of the conditional probabilities from Eqs.~(\ref{P10-01}) as
\begin{subequations}
  \begin{eqnarray}
  \langle n_{0,k}^1 \rangle &=& k P(1|0) = \frac{k \rho}{2(1-\sigma)}
  ~~~ \mbox{and} \\
  ~~~ \langle n_{1,k}^0 \rangle &=& k P(0|1) = \frac{k \rho}{2 \sigma}.
\end{eqnarray}
\label{1st}
\end{subequations}
Finally, replacing these expressions for the first moments in
Eq.~(\ref{ds-dt-1}) and rearranging terms we arrive at the
following equation for the evolution of the fraction of perfect agents quoted in Eq.~(\ref{dsdt-nets}) of the main text:
\begin{eqnarray}
    \frac{d\sigma}{dt} &=& \frac{\rho}{2} \left[ p (P_{1|0}-P_{0|1}) +
                      (1-p)(\tilde{P}_{1|0} - \tilde{P}_{0|1})
                      (1-\mathcal{P}_1) \right],
\end{eqnarray}
where the payoffs $P$ are given by Eqs.~(\ref{P01}), (\ref{P10}) and (\ref{s01-10-1}).

\subsection{Equation for $\rho$}
\label{Eq-rho}

To derive an equation for the fraction of active links $\rho$ we follow
an approach similar to that of Section~\ref{Eq-s} for $\sigma$.  In Fig.~\ref{update}(a) we
describe the change in $\rho$ in a time step $\Delta t=1/N$, when a node $i$ with state
$\theta_i=0$ and degree $k$ copies the state $\theta_j=1$ of a random
neighbor $j$ ($01 \to 11$ transition).  The $n_{0,k}^1$ active links
($0$--$1$) connected to node $i$ become inert ($1$--$1$) and
viceversa, leading to a local change $\Delta n_{0,k}^1 = k-2
n_{0,k}^1$ in the number of active links and a change 
$\Delta \rho = \frac{2(k-2 n_{0,k}^1)}{\mu N}$ in $\rho$, where $\mu
N/2$ is the total number of links in the system.  Similarly, in a $10
\to 00$ transition the change is $\Delta \rho = \frac{2(k-2
  n_{1,k}^0)}{\mu N}$.  Considering the two transitions and the 
two playing strategies, we can write the following equation analogous
to Eq.~(\ref{ds-dt}):
\begin{eqnarray}
	\frac{d\rho}{dt} &=& \frac{(1-\sigma)}{1/N} \Bigg\{ p
                             \sum_{k=1}^{N-1} \mathcal{P}_k
                             \sum_{n_{0,k}^1=0}^k B \left( n_{0,k}^1
                             \right) \frac{n_{0,k}^1}{k}  \, P_{1|0} \,
                             \frac{2 \left( k-2 n_{0,k}^1 \right)}{\mu
                             N} 
                             + (1-p) \sum_{k=2}^{N-1} \mathcal{P}_k
                             \sum_{n_{0,k}^1=0}^k B \left( n_{0,k}^1
                             \right) \frac{n_{0,k}^1}{k} \,
                             \tilde{P}_{1|0} \, \frac{2 \left( k-2
                             n_{0,k}^1 \right)}{\mu N} \Bigg\}  \nonumber \\
	&+& \frac{\sigma}{1/N} \Bigg\{ p \sum_{k=1}^{N-1} \mathcal{P}_k
            \sum_{n_{1,k}^0=0}^k B \left( n_{1,k}^0 \right)
            \frac{n_{1,k}^0}{k}  \, P_{0|1} \, \frac{2 \left( k-2
            n_{1,k}^0 \right)}{\mu N}
            + (1-p) \sum_{k=2}^{N-1} \mathcal{P}_k
                \sum_{n_{1,k}^0=0}^k B \left( n_{1,k}^0 \right)
                \frac{n_{1,k}^0}{k} \, \tilde{P}_{0|1} \, \frac{2 \left( k-2
            n_{1,k}^0 \right)}{\mu N} \Bigg\}.
\label{drho-dt}
\end{eqnarray}
As we have done in Section~\ref{Eq-s}, we can pull $P_{1|0}$,
$\tilde{P}_{1|0}$, $P_{0|1}$ and $\tilde{P}_{0|1}$ out of the
summations, combine the first and second summations over $k$ into one,
as well as the third and fourth summations, and perform the summations
over the number of active links connected to node $i$, to obtain
\begin{eqnarray}
  \frac{d\sigma}{dt} &=& \frac{2(1-\sigma)}{\mu} \Bigg\{ \left[ p P_{1|0} + (1-p)
                          \tilde{P}_{1|0} \right] \sum_{k=1}^{N-1}
                          \frac{\mathcal{P}_k}{k} \left[ k \langle
                    n_{0,k}^1 \rangle - 2 \langle (n_{0,k}^1)^2 
                    \rangle \right] - (1-p) \tilde{P}_{1|0} \mathcal{P}_1
                         \left( k \langle n_{0,1}^1 \rangle - 2
                    \langle ( n_{0,1}^1 )^2 \rangle \right) \Bigg\}  \nonumber \\ 
  &+& \frac{2 \sigma}{\mu} \Bigg\{ \left[ p P_{0|1} + (1-p)
                          \tilde{P}_{0|1}  \right] \sum_{k=1}^{N-1}
                          \frac{\mathcal{P}_k}{k}  \left[
      k \langle n_{1,k}^0 \rangle - 2  \langle (n_{1,k}^0)^2 \rangle \right]  -
                          (1-p) \tilde{P}_{0|1} \mathcal{P}_1
                          \left[ \langle n_{1,1}^0 \rangle -  2
      \langle (n_{1,1}^0)^2 \rangle \right] \Bigg\}.
      \label{drho-dt-1}
\end{eqnarray}
Here $\langle (n_{0,k}^1)^2 \rangle$ and $\langle (n_{1,k}^0)^2 \rangle$ are
the second moments of the Binomial distributions $B \left(n_{0,k}^1
\right)$ and $B \left( n_{1,k}^0 \right)$, respectively, which are
given by
\begin{subequations}
\begin{eqnarray}
  \langle (n_{0,k}^1)^2 \rangle &=& k P(1|0) + k(k-1) P(1|0)^2 =
                 \frac{k \rho}{2(1-\sigma)} + \frac{k(k-1) \rho^2}{4(1-\sigma)^2}
  ~~~ \mbox{and} \\
  ~~~ \langle (n_{1,k}^0)^2 \rangle &=& k P(0|1) + k(k-1) P(0|1)^2
        = \frac{k \rho}{2 \sigma} + \frac{k(k-1) \rho^2}{4 \sigma^2}.
\end{eqnarray}
\label{2nd}
\end{subequations}
Finally, plugging these expressions for the second moments and the expressions
from Eqs.~(\ref{1st}) for the first moments into
Eq.~(\ref{drho-dt-1}) we obtain, after doing some algebra, the
following equation for the evolution of the fraction of active links:
\begin{eqnarray*}
\frac{d\rho}{dt} &=& \frac{\rho}{\mu} \Bigg\{ \left[ p P_{0|1} +
                         (1-p) \tilde{P}_{0|1} \right] \left[ (\mu-1)
                         \left(1-\frac{\rho}{\sigma} \right) - 1 \right] +
                         \left[ p P_{1|0} + (1-p) \tilde{P}_{1|0}
                         \right] \left[ (\mu-1) \left(
                         1-\frac{\rho}{1-\sigma} \right) - 1 \right] \\
    &+& (1-p) \left( \tilde{P}_{0|1} + \tilde{P}_{1|0} \right) \mathcal{P}_1 \Bigg\},
\end{eqnarray*}
which is quoted in Eq.~(\ref{drhodt}) of the main text.

\end{widetext}

\section*{References}

\bibliography{references}

\end{document}